\documentclass[%
 reprint,
superscriptaddress,
 amsmath,amssymb,
 aps,
]{revtex4-2}

\usepackage{graphicx}
\usepackage{dcolumn}
\usepackage{bm}
\usepackage{braket}
\usepackage{color}
\usepackage{amsthm}
\usepackage{dsfont}
\usepackage[absolute]{textpos}
\usepackage{xcolor}
\usepackage{fancyhdr}
\usepackage{natbib}
\makeatletter
\let\cat@comma@active\@empty
\makeatother

\usepackage{appendix}

\begin{document}

\preprint{APS/123-QED}

\title{Generation of polarization entanglement via the quantum Zeno effect}

\author{Ian C. Nodurft}
\email[]{inodurft@uic.edu}
\affiliation{Dept.~of Electrical \& Computer Engineering, University of Illinois Chicago, Chicago, IL 60607, USA}
\affiliation{Peraton, Greenbelt, MD 20770, USA}
\author{Harry Shaw}
\affiliation{NASA Goddard Space Flight Center, Greenbelt, Maryland 20771, USA}
\author{Ryan~T. Glasser}
\affiliation{Department of Physics and Engineering Physics, Tulane University, New Orleans, Louisiana 70118, USA}
\author{Brian~T. Kirby}
\affiliation{Department of Physics and Engineering Physics, Tulane University, New Orleans, Louisiana 70118, USA}
\affiliation{DEVCOM US Army Research Laboratory, Adelphi, MD 20783, USA}
\author{Thomas~A. Searles}
\email[]{tsearles@uic.edu}
\affiliation{Dept.~of Electrical \& Computer Engineering, University of Illinois Chicago, Chicago, IL 60607, USA}

\date{\today}

\begin{abstract}
The quantum Zeno effect reveals that the continuous observation of a quantum system can result in significant alterations to its evolution. Here, we present a method for establishing polarization entanglement between two initially unentangled photons in coupled waveguides via the quantum Zeno effect. We support our analytical investigation with numerical simulations of the underlying Schrodinger equation describing the system. Further, we extend our technique to three coupled waveguides in a planar configuration and determine the parameter regime required to generate three-qubit W-states. Our findings offer a powerful quantum state engineering approach for photonic quantum information technologies. 
\end{abstract}

\maketitle

\section{Introduction}

The quantum Zeno effect (QZE) is a phenomenon whereby repeated projective measurements can slow or halt the temporal evolution of a quantum system \cite{misra1977zeno,itano1990quantum,kofman1996quantum,schulman1998continuous}. First introduced in the context of an apparent paradox in the decay times of particles observed in bubble chambers \cite{misra1977zeno}, the QZE has since been experimentally verified using various platforms, including two-level \cite{itano1990quantum}, multi-level \cite{Schfer2014}, and superconducting systems \cite{matsuzaki2010quantum}, and even on quantum computers \cite{barik2020demonstrating}, among others. From a theoretical perspective, the QZE has also come to encompass a more generalized set of phenomena. These include the quantum anti-Zeno effect \cite{kofman2000acceleration,kofman2001zeno}, where repeated observation can accelerate particle decay, and quantum Zeno dynamics \cite{facchi2008quantum,facchi2001zeno,facchi2002quantum,smerzi2012zeno}, where frequent measurement does not confine a system to a single state but rather a multidimensional subspace in which evolution can still occur.

Aside from the foundational importance of the QZE in understanding the role of measurement in quantum mechanics, it has also matured into a practical tool for developing quantum information technology. For example, the QZE has been used to realize quantum gates \cite{franson2004quantum,you2012theoretical}, decoherence-free subspaces \cite{beige2000quantum}, and quantum error correction \cite{paz2012zeno,erez2004correcting,dominy2013analysis,chen2020continuous}. 
Additionally, the QZE can generate entanglement, making it a potentially valuable tool for state preparation in quantum systems.
While implementation agnostic procedures for generating entanglement via the QZE exist \cite{wang2008quantum, magazzu2018generation,nakazato2004preparation}, platform-specific proposals have also been developed for experimental setups involving optical cavities \cite{li2017engineering,shao2010converting}, spatially separated cavities \cite{li2012controllable,liu2013achieving}, atomic ensembles \cite{barontini2015deterministic}, and photonic systems \cite{ten2011entangling}. In the latter case, path-entangled photons have been considered where the role of the QZE is to suppress the transition of a target photon from one optical mode to another based on the presence of a control photon. However, in many practical applications of entanglement, path-entangled photons present a challenge due to the necessity of high detector efficiencies since the presence of zero photons is a valid outcome \cite{Knill2001,nunn2021heralding}.

In this work, we consider a method for creating two- and three-photon polarization-entangled states using coupled waveguides mediated by the QZE. In our approach, the QZE is used to suppress unwanted bunching terms analogous to the proposal of Franson \cite{franson2004quantum,you2012theoretical}. However, novel to our approach, we extend the system to polarization encoded qubits by proposing the use of a diamond-type four-level system to mediate the QZE. Although the QZE suppresses the bunching of the inserted photons at the output, it does not inhibit the exchange of photon positions in the waveguides, assuming such transitions can occur. Hence, by engineering the waveguide length and coupling, it is possible to create uncertainty as to which output mode the inserted photons exit, resulting in an entangled state. 

\begin{figure*}[ht!]
    \centering
    \centerline{\includegraphics[scale=1]{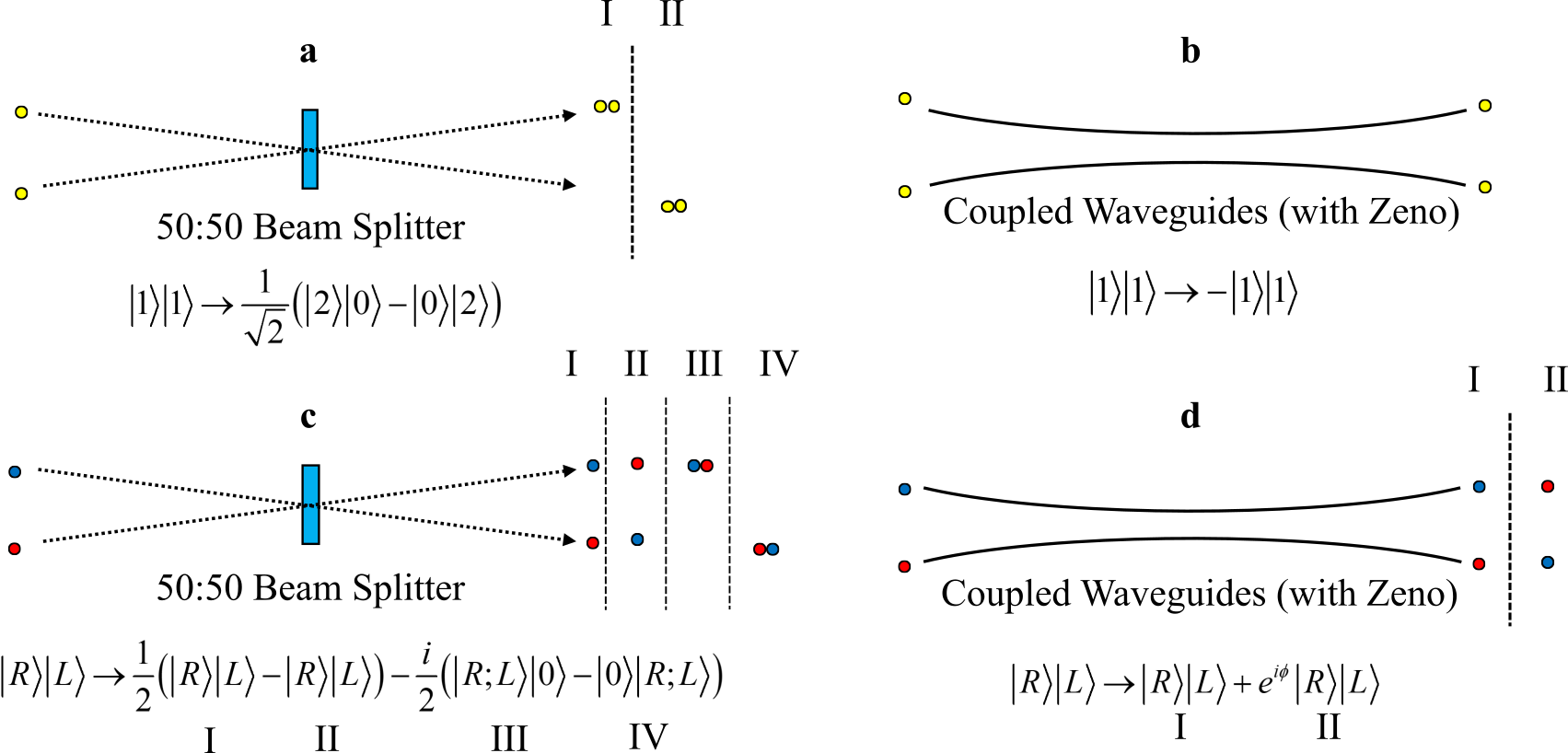}}
    \caption{Comparison of various devices acting on two input photons.  In all cases, one photon is injected into each of the two modes, with yellow indicating identical photons and red and blue indicating photons with orthogonal polarizations.  a)  Two identical and unentangled photons incident on a 50:50 beam splitter.  Photon bunching occurs and both photons exit in the same mode, though which mode is not determined.  b)  Identical photons incident on the device proposed by \cite{franson2004quantum}.  An observing medium (3-level ladder type atoms) suppresses the bunching effect present in a), producing a nonlinear phase-shift in the quantum state.  c)  One photon each of right and left circular polarization incident on a 50:50 beam splitter.  Bunching is still present, but due to the distinguishable nature of the photons the anti-bunched states are still present.  d)  One photon each of right and left circular polarization injected into a very similar device as b).  Bunching is suppressed, leaving the state in a superposition of the two anti-bunched states, yielding polarization entanglement.}
    \label{fig:BSFransonComp}
\end{figure*}

The paper is organized as follows. We begin by providing a quantum mechanical model of our system and perform analytical calculations based on this model that demonstrate how the QZE results in the suppression of photon bunching. We then perform a numerical simulation of this physical system and show that initially unentangled input photons can evolve into two-photon Bell states and three-photon W states. Finally, we conclude with a discussion and analysis of our results.

\section{Theoretical Model}

The overall working principle of our polarization entanglement generation scheme relies on the established ability of the QZE to suppress photon bunching at a beamsplitter. When an unentangled pair of indistinguishable single photons is incident on a balanced beamsplitter with one photon per input port, the output state includes bunched terms where both photons exit the same port as visualized in Fig.~\ref{fig:BSFransonComp}a. This bunching effect is considered a failure mode for linear quantum computing schemes \cite{Knill2001}. Previously, Franson showed that the QZE could effectively suppress these bunching terms in mode entangled systems \cite{franson2004quantum,you2012theoretical}. 
In particular, Franson showed that the QZE could cause input photons to act as fermions in this scenario and proposed a technique for realizing a controlled-phase (CZ) gate by exploiting this effect. We depict the original CZ gate of Franson in Fig.~\ref{fig:BSFransonComp}b, where all bunching terms are suppressed in contrast to the beamsplitter configuration of Fig.~\ref{fig:BSFransonComp}a.
Note that we replace the beamsplitter from Fig.~\ref{fig:BSFransonComp}a with a coupled waveguide system in Fig.~\ref{fig:BSFransonComp}b. Coupled waveguides can implement beamsplitter transformations \cite{makarov2020theory} and allow for direct implementation of the QZE using a series of atoms that observe the system as it evolves \cite{franson2004quantum}.

Ultimately, the bunching and anti-bunching behavior at the beamsplitter (also observed in waveguide systems \cite{makarov2020theory}) in Figs.~\ref{fig:BSFransonComp}a,b is due to the indistinguishability of the input photons.
If we also consider the incoming photon polarization, the behavior can become more complicated.
For example, in Fig.~\ref{fig:BSFransonComp}c, a balanced beamsplitter acts on one right and one left circularly polarized photon. Unlike Fig.~\ref{fig:BSFransonComp}a, the output state now includes two anti-bunched terms. Our approach (depicted in Fig.~\ref{fig:BSFransonComp}d) allows for the situation where two unentangled photons with different polarizations (in this case, one right circularly polarized and one left circularly polarized) evolve into an entangled state through suppression of the bunching terms shown in Fig.~\ref{fig:BSFransonComp}c. Note that this method is in some ways analogous to Fig.~\ref{fig:BSFransonComp}b, but importantly our method allows for the treatment of input photons with different polarizations. We exploit the suppression of photon bunching to restrict the evolution of the coupled waveguide system to only include states where each photon exits a different port and engineer the coupling parameters of the system so that the two terms are equally probable. 

Although the original CZ gate proposed in \cite{franson2004quantum} can generate mode-entangled photons, 
polarization-entangled photons as generated by the technique in Fig.~\ref{fig:BSFransonComp}d have several practical advantages. 
In particular, using the vacuum mode as an information carrier necessitates high-efficiency detectors for implementation \cite{Knill2001}.
Additionally, since the absence of a photon is a valid basis for the state generated in \cite{franson2004quantum}, the action of pure-loss can directly impact entanglement quality \cite{sangouard2007long}. 
Alternatively, our method produces entanglement in the polarization degree of freedom which decouples pure photon loss from the encoding basis and can be manipulated directly using waveplates that are easily accessible experimentally \cite{crespi2011integrated}.
Furthermore, as we will see, our technique can be applied to transformations of more than two qubits. We can modify the configuration by adding an additional waveguide and placing them in a planar configuration (shown in Fig. \ref{fig:ZenoW}), which allows for the generation of W-states with our scheme \cite{PhysRevA.62.062314}.

\begin{figure}[b!]
    \centering
    \includegraphics[width=0.45\textwidth]{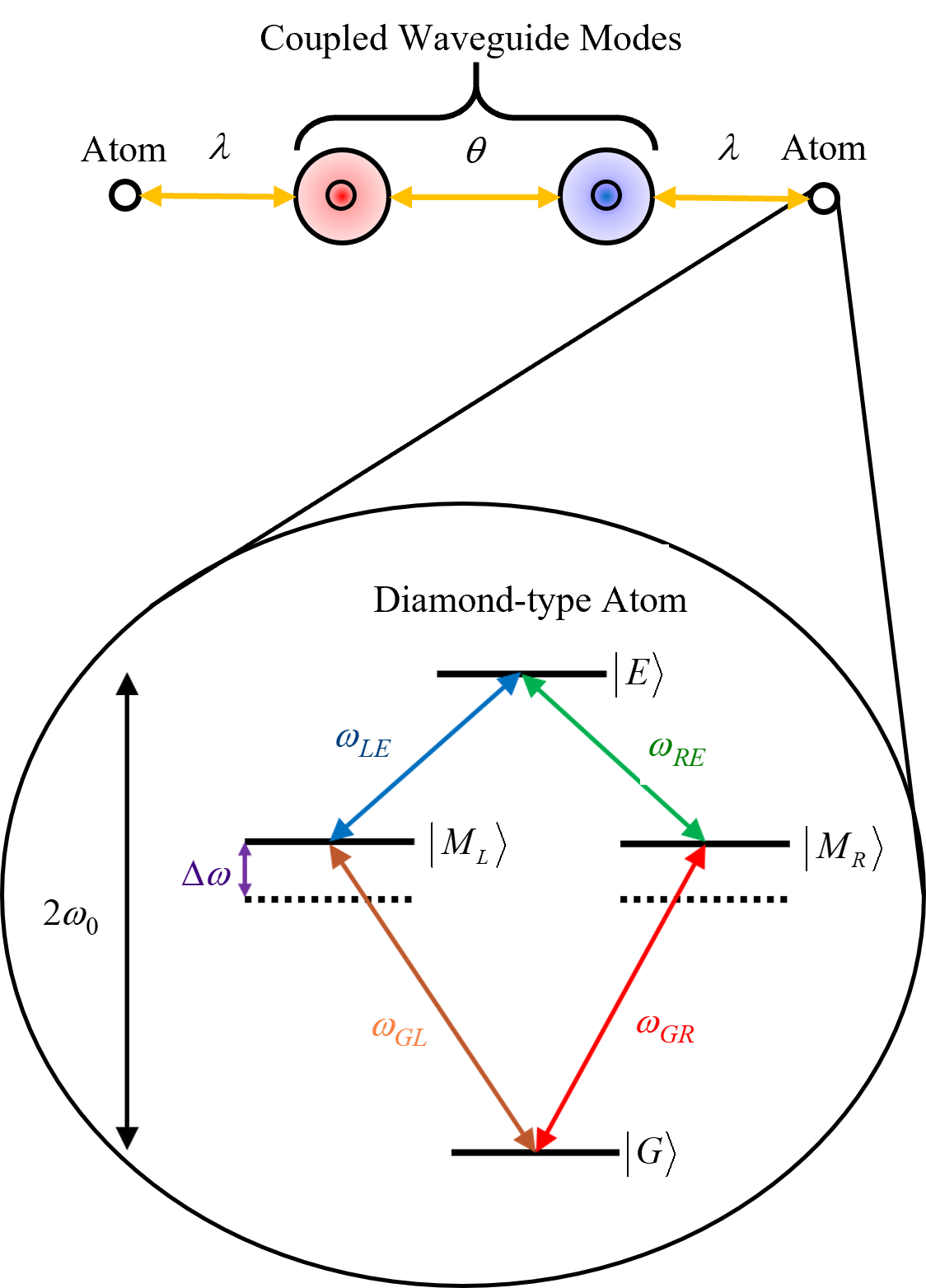}
    \caption{Displayed at the top is the physical design of the entangling Zeno gate. A pair of waveguides are coupled together with strength $\theta$ and additionally to an absorbing medium, such as an atom. The diagram of a 4-level atom in the diamond configuration is displayed on the bottom. The middle levels $| M_L \rangle$ and $| M_R \rangle$ are detuned from resonance with a single photon, and the excited state $| E \rangle$ is on resonance with a pair of photons.}
    \label{fig:FourLevel}
\end{figure}

To realize the QZE in our system, we must choose an observing medium that can absorb photon pairs, destroying any bunched states.  Furthermore, it must not experience significant single-photon absorption so that those states remain unaffected.
Here, we will assume the input photons are circularly polarized with one right and one left.
Hence, the angular momentum of the excited state must be the same as that of the ground state.  Ideally, we also need intermediate states that are sensitive to either right or left polarization, such that their angular momentum differs from the ground, and consequently the excited state, by $\pm 1$.  Naturally, this leads us to the diamond configuration shown in Fig. \ref{fig:FourLevel}.
The diamond-configuration has four unique energy levels: a ground state $\left| G \right\rangle $, excited state $\left| E \right\rangle $, and two middle energy levels $\left| M_L \right\rangle $ and $\left| M_R \right\rangle $.  By our convention, $\left| M_L \right\rangle $ and $\left| M_R \right\rangle $ are named based on the difference in angular momentum between themselves and the ground state.  Left circularly polarized photons, therefore, facilitate transitions from $\left | G \right \rangle$ to $\left| M_L \right\rangle $ and right circularly polarized photons facilitate transitions between $\left | G	\right \rangle$ and $\left| M_R \right\rangle$.

In our investigation, we assume that incoming photons have an energy of ${{E}_{R}}={{E}_{L}}=\hbar \omega $.  The energy of the absorber ground state is ${{E}_{G}}=0$, the middle levels are ${{E}_{M_R}}=\hbar (\omega _0+\Delta_R)$ and ${{E}_{M_L}}=\hbar (\omega _0+\Delta_L)$, and the excited state is ${{E}_{E}}=2{{\hbar\omega }_{0}}$.  We assume that ${{\omega }_{0}}=\omega $ so that the effective transition from ground state to excited state is on resonance with a pair of photons.  The middle levels are detuned from the single-photon resonance by $\Delta_{R}=\Delta_{L}=\Delta $ where $\Delta$ is large enough such that single-photon absorption rates are relatively slow compared to two-photon absorption rates.  In Fig.~\ref{fig:FourLevel}, this means that ${{\omega }_{GL}}={{\omega }_{GR}}=\omega +\Delta $ and ${{\omega }_{LE}}={{\omega }_{RE}}=\omega -\Delta$.  In conjunction with the assumption of strong atomic coupling, this facilitates strong two-photon absorption with negligible absorption of single-photons of either polarization.

Our proposed system is pictured at the top of Fig. \ref{fig:FourLevel} where a pair of waveguides are coupled together and to an absorbing medium, such as an atom. The Hamiltonian for this system is $\hat{H} (t) = \hat{H}_0 +\hat{H}_I$, where $\hat{H}_0$ is the rest Hamiltonian and $\hat{H}_I$ is the interaction Hamiltonian.  They are given by:

\begin{equation}
    \hat{H}_0 = \hbar \omega (\hat{a}_R^\dagger \hat{a}_R + \hat{a}_L^\dagger \hat{a}_L + \hat{b}_R^\dagger \hat{b}_R + \hat{b}_L^\dagger \hat{b}_L) +\hbar \omega_0 (\hat{A}_3 + \hat{B}_3)
    \label{eq:RestHam}
\end{equation}

and

\begin{equation}
   \begin{split}
        \hat{H}_I (t) &= \hbar \theta (\hat{a}_R^\dagger \hat{b}_R + \hat{b}_R^\dagger \hat{a}_R + \hat{a}_L^\dagger \hat{b}_L + \hat{b}_L^\dagger \hat{a}_L)\\
        & + \hbar \lambda (\hat{A}_{GR} \hat{a}_R + \hat{A}_{LE} \hat{a}_R + \hat{A}_{GL} \hat{a}_L + \hat{A}_{RE} \hat{a}_L)\\
        & + \hbar \lambda (\hat{A}_{RG} \hat{a}^\dagger_R + \hat{A}_{EL} \hat{a}^\dagger_R + \hat{A}_{LG} \hat{a}^\dagger_L + \hat{A}_{ER} \hat{a}^\dagger_L)\\
        & + \hbar \lambda (\hat{B}_{GR} \hat{b}_R + \hat{B}_{LE} \hat{b}_R + \hat{B}_{GL} \hat{b}_L + \hat{B}_{RE} \hat{b}_L)\\
        & + \hbar \lambda (\hat{B}_{RG} \hat{b}^\dagger_R + \hat{B}_{EL} \hat{b}^\dagger_R + \hat{B}_{LG} \hat{b}^\dagger_L + \hat{B}_{ER} \hat{b}^\dagger_L)
   \end{split}
   \label{eq:IntHam}
\end{equation}

where $\theta$ is the coupling between waveguide modes and $\lambda$ is the coupling between each waveguide and their respective atom, as in Fig.~\ref{fig:FourLevel}.  Operators $\hat{A}_3$ and $\hat{B}_3$ define the energy levels of the atom, and $\hat{A}_{GR}$/$\hat{B}_{GR}$, $\hat{A}_{GL}$/$\hat{B}_{GL}$, $\hat{A}_{LE}$/$\hat{B}_{LE}$, and $\hat{A}_{RE}$/$\hat{B}_{RE}$ facilitate transitions between each of these levels.  The operators $\hat{a}_R$, $\hat{a}_L$, $\hat{b}_R$, and $\hat{b}_L$ and their adjoints are the standard ladder operators for right and left polarized photons with the $a$ operators referring to one waveguide and the $b$ the other.  Transforming to the interaction picture, Eqn.~(\ref{eq:IntHam}) has complex exponentials of the form $\text{exp}(\pm \Delta t)$ attached to all terms in its 2nd and 3rd lines due to the detuning of the intermediate levels.

To determine the transition rates for photons interacting with the diamond-type atoms pictured in Fig.~\ref{fig:FourLevel}, we can treat the interaction Hamiltonian as a perturbation and ultimately use Fermi's Golden Rule \cite{baym1973lectures}.  In particular, we will consider the first and second-order transitions of the diamond-type absorber.  We assume the absorbers always begin in the ground state, first-order transitions will include transitions from the ground state $| G \rangle$ to either the right $| M_R \rangle$ or left $| M_L \rangle$ intermediate state, and second-order transitions are those from the ground $| G \rangle$ to the excited state $| E \rangle$.  Both the first and second order transition probabilities as well as the corresponding transition rates can be calculated using time-dependent perturbation theory, the adiabatic approximation, and Fermi's first and second order Golden Rules \cite{baym1973lectures}.  These are well known concepts so we will omit them here, but we present the derivation in Appendix \ref{app:TDPT} for clarity. 

We begin by defining  the atomic levels to be

\begin{equation}
    \begin{split}
        &E_G = 0,\\
        &E_{M_R} = E_{M_L} = \hbar (\omega + \Delta),\\
        &E_E = 2 \hbar \omega.
    \end{split}
    \label{eq:EnergyLevels}
\end{equation}
Here, $\omega$ is the frequency of the incident photons, giving them energy $E_p = \hbar \omega$, and $\Delta$ is the detuning from the photon by the atom resonant frequency.  For single-photon absorption, $E_{M_i} - E_p = \hbar \Delta$ and for two photon absorption, $E_E - 2E_p = 0$. Plugging Eqns.~(\ref{eq:IntHam}) and (\ref{eq:EnergyLevels}) into the first and second Golden Rules yields the transition rates

\begin{equation}
    \begin{split}
        \Gamma^{(1)}_{G \to M_i} &= \frac{2 \pi}{\hbar} (\hbar \lambda)^2 \delta (\Delta \hbar \omega)\\
        &= 0
    \end{split}
    \label{eq:OneAbsorb}
\end{equation}

\begin{equation}
    \begin{split}
        \Gamma_{G \to E}^{(2)} &= \frac{2t}{\hbar^2} \frac{4 [\hbar \lambda]^4}{(\Delta \hbar \omega)^2}\\
        &= \frac{8 t \lambda^4}{(\Delta \omega)^2}
    \end{split}
    \label{eq:TwoAbsorb}
\end{equation}

where the superscripts $(1)$ and $(2)$ label the first-order and second-order transitions, and $M_i$ can be either middle absorber level.  Comparing the results of Eqns.~(\ref{eq:OneAbsorb}) and (\ref{eq:TwoAbsorb}), it is clear that single-photon absorption rates are much smaller than two-photon absorption rates.  Thus, we can expect that single photons will be unimpeded by the presence of the absorbers, but a photon pair will not.
This imbalance between single and two-photon absorption rates facilitates the Zeno effect by suppressing terms where both a left and right photon are found, but not when a single photon is found.  Intuitively, this means we eliminate the bunching effect throughout the process and should be able to achieve the transformation shown in Fig. \ref{fig:BSFransonComp}d.

We now consider how the two-photon absorption process occurs in the physical scenario described in Figs. \ref{fig:BSFransonComp}d and \ref{fig:FourLevel}.  Since we do not start in a bunched state, the two-photon absorption is facilitated by a third-order process where one photon switches waveguide modes and consequently both photons become absorbed.  The rate at which this process occurs must remain faster than the single-photon absorption if we are to fully suppress the bunching effect.  To that end, we calculate three probability amplitudes; the first case where the initial state (i.e., one photon in either waveguide) transitions to a bunched state, a second case where each photon switches waveguide modes, and the third case where one photon moves to the opposite mode and then this bunched pair is absorbed. These three probability amplitudes correspond to a first, second, and third-order transition. Up to nth order, the transition probability amplitudes between waveguide modes are again found with time-dependent perturbation theory \cite{baym1973lectures}. 
The only nonzero matrix elements for odd-order transitions are those with  bunched photons, while for even-order transitions the only nonzero matrix elements are those that will evolve as anti-bunched.  As such, every $n$-th order transition follows a noticeable trend.  These matrix elements are given by

\begin{equation}
    \langle f | \psi (t) \rangle^{(n)} = \frac{1}{n!} \frac{2^{n-1} (\theta t)^n}{i^n}.
    \label{eq:WaveguideTrans}
\end{equation}

Assuming an initial photon state $| \psi_p (0) \rangle = | R \rangle_a | L \rangle_b$, the first-order transition to either bunched state has transition probability amplitude:

\begin{equation}
    p_{B}=\langle f | \psi (t) \rangle^{(1)} = -i \theta t,
    \label{eq:WaveguideFirst}
\end{equation}

and the second-order transition probability amplitude to the alternative anti-bunched state $| L \rangle_a | R \rangle_b$ is:

\begin{equation}
   p_{AB}= \langle f | \psi (t) \rangle^{(2)} = - \theta^2 t^2.
    \label{eq:WaveguideSecond}
\end{equation}

Finally, transitions to the excited state of either atom are third-order, requiring one photon to switch waveguides and then both to become absorbed.  Either transition has the probability amplitude:

\begin{equation}
    p_{TPA}=\langle f | \psi (t) \rangle^{(3)} = \frac{3 \theta \lambda^2 t^2}{2 \Delta} + \frac{3 i \theta \lambda^2 t}{\Delta^2} + \frac{3 \theta \lambda^2}{\Delta^3} \Big ( e^{-i \Delta t} - 1 \Big ).
     \label{eq:ThirdOrder}
\end{equation}

Each of these probability amplitudes corresponds to a probability given by:

\begin{equation}
    P_{B}=\left| p_{B} \right|^2, 
    \label{eq:ProbWaveFirst}
\end{equation}

\begin{equation}
    P_{AB}=P_{B}=\left| p_{AB} \right|^2,
    \label{eq:ProbWaveSecond}
\end{equation}

and

\begin{equation}
    P_{TPA}=P_{B}=\left| p_{TPA} \right|^2.
    \label{eq:ProbThird}
\end{equation}

The third-order process is dominated by different terms depending on the time-scale over which the process takes place.  For $t \to 0$, the middle term $3 i \theta \lambda^2 t / \Delta^2$ dominates and for $t \to \infty$ the first term $3 \theta \lambda^2 t^2 / 2 \Delta$ dominates.
We need only concern ourselves with the short time, as the Zeno effect can be interpreted as the result of rapid measurements \cite{misra1977zeno}.

We now compare $p_{B}$, $p_{AB}$, and $p_{TPA}$ in the short time regime.
Both $p_{B}$ and $p_{TPA}$ are proportional to $t^2$ while $p_{AB}$ is proportional to $t$ and hence when $t<<1$ the $p_{AB}$ amplitude is dominated by both $p_{B}$ and $p_{TPA}$.
However, the third-order process leading to $p_{TPA}$ we have described here is mutually exclusive to the second-order process leading to $p_{AB}$, so the two-photon absorption does not impede this particular transition. 
Furthermore, due to significant detuning, the single-photon absorption process is much slower than the transitions leading to anti-bunching, as suggested by $\Gamma^{(1)}_{G \to M_i}$.  Therefore, we expect that the single-photon absorption process will not destroy the state quickly enough to stop the transition.

In addition, under the assumption of strong atomic coupling, i.e. $\lambda >> \theta$, and with non-negligible detuning $\Delta$, the two-photon absorption rate becomes much larger than the single-photon transition (bunching) rate.  Each transition rate, as per the golden rule, is found by taking the time-derivative of $P_{B}$, $P_{AB}$, and $P_{TPA}$.  In keeping with the notation of Eqn. (\ref{eq:OneAbsorb}) and (\ref{eq:TwoAbsorb}), these transition rates are given by $\Gamma_B$, $\Gamma_{AB}$, and $\Gamma_{TPA}$ respectively.  The single-photon waveguide transition, or bunching, rate is $\Gamma_{B}\rightarrow~\theta^2$, irrespective of the strength of atomic coupling or whether we are in the short or long time regime.  By the same method, the third-order transition facilitating two-photon absorption, Eqn. (\ref{eq:ProbThird}), is found to proceed at the rate $\Gamma_{TPA}\rightarrow~\theta^2 \lambda^4$ in the short time regime. 

Comparing these two rates, we see that in the strong atomic coupling limit of $\lambda >>\theta$, the probability of two-photon absorption grows more rapidly than the probability of the system transitioning to a bunched configuration.

Further, despite the two-photon transition probability being more significant in a relative sense than the bunching probability, both transitions are overall small quantities and thus not the dominant terms of the total system evolution. Hence, over short time scales, the effective measurements made by the two-photon absorption process keep the state evolving primarily through the anti-bunching transition.

In conclusion, in the $t\rightarrow 0$ limit imposed by the QZE, the two-photon absorption and bunching processes are comparatively dominant to the anti-bunching process. Additionally, by assuming strong atomic coupling where $\lambda>>\theta$, we can make the two-photon absorption rate more significant than the rate of bunching, therefore suppressing the bunching terms. When these limits are considered together along with the relatively low single-photon absorption rate, found from $\Gamma^{(1)}_{G\rightarrow M_{i}}$ we see the dominant remaining process in the system to be anti-bunching.
Hence, we expect the system will oscillate between the two anti-bunched terms where these two photons exchange positions. Accordingly, at one or more stages of the process, we should have an equal probability of finding the system in either of the two anti-bunched states. The existence of these oscillations means that correctly choosing the coupling or length of the coupled waveguide region will reproduce the expected result shown in Fig. \ref{fig:BSFransonComp}d.

\section{Numerical results and discussions}

In the previous sections, we have argued based on calculated transition probabilities and projective probability amplitudes that we expect the QZE to effectively suppress bunching terms in our system.  However, the suppression of bunched terms alone will not necessarily result in highly entangled states and rather the detailed configuration of the coupled waveguides will ultimately determine the characteristics of the output state.  In this section we perform full numerical simulations of the underlying Schrodinger equation describing our system and extract various quality metrics to characterize the system.  Further, we apply these techniques to systems of three coupled waveguides in a planar configuration with three input photons and numerically simulate the evolution of this system to demonstrate the creation of W states.

\begin{figure}[ht]
    \centering
    \includegraphics[width=0.45\textwidth]{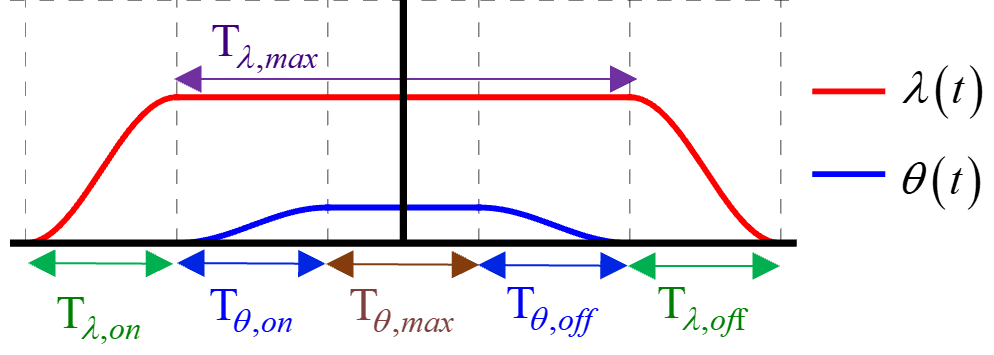}
    \caption{The time dependence of the coupling parameter between waveguide modes $\theta (t)$ and atomic coupling $\lambda (t)$.  The atomic coupling is turned on for a time $T_{\lambda,on}$ eventually reaching a maximum at which point the waveguide coupling follows suit for time $T_{\theta,on}$.  After maintaining a maximum for $T_{\theta,max}$ and $T_{\lambda,max}$, they are adiabatically turned off in reverse order.}
    \label{fig:Adiabatic}
\end{figure}

In our simulation, $\theta$ and $\lambda$ are time-dependent as depicted in Fig. \ref{fig:Adiabatic} and are defined such that the adiabatic approximation applies.  In order to suppress photon-bunching, we need to begin measuring the photon state before the photons can begin switching waveguides.  For this reason, the atom-waveguide coupling, $\lambda$, is "turned on" first.  The adiabatic approximation helps avoid spurious excitations and mode swapping.  Furthermore, for simplicity, the simulation only uses one absorber in each waveguide mode rather than many; however, because the absorber coupling is "on" for the entire interaction period, it simulates a state of continuous observation and, as such, approximates the situation of many observers configured to maintain observation for the entire evolution.  Alternatively, from the projective measurement point of view, it is effectively the case where the number of projective measurements we perform tends to infinity.

\begin{figure}[ht!]
    \centering
    \includegraphics[width=0.45\textwidth]{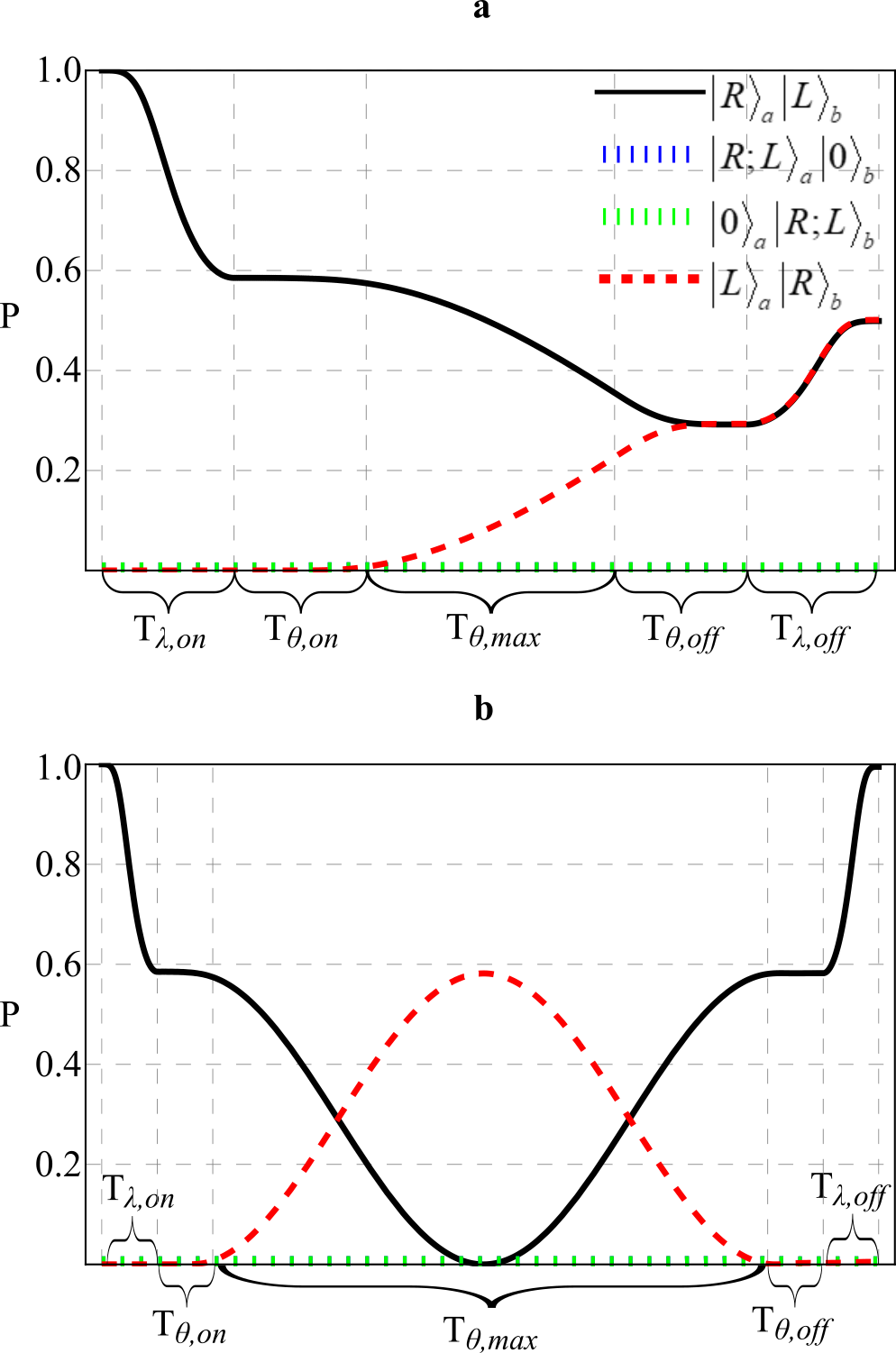}
    \caption{The time-evolution of those four states where both photons remain in the two waveguide modes.  The black (solid) curve is the initial state $| \psi \rangle = | R \rangle_a | L \rangle_b | G \rangle_A | G \rangle_B$, and the red (dashed) curve is the state where both photons have flipped waveguides.  The blue and green (both dotted) curves are those two states where the photons are in the same waveguide mode.  a)  The red curve gradually increases until it meets with the blue curve so the anti-bunched states have the same probability, the interactions are turned off yielding 50\% probability for both cases.  b)  An extended interaction under the same conditions.  Rabi oscillations appear between the two anti-bunched states.}
    \label{fig:ForcedEntanglement}
\end{figure}

Preparing all relevant parameters and builiding the Hamiltonian, we simulate the system using the time-dependent Schrodinger equation.  Combining this with the initial condition that $\vert\psi (t_{in}) \rangle = | R \rangle_a | L \rangle_b | G \rangle_A | G \rangle_B$, where $| X \rangle_i$ denotes an $X$ polarized photon in mode $i$, we numerically solve for $| \psi (t) \rangle$.  Bunched states are similarly denoted by $| X;Y \rangle_i$ as seen in Fig. \ref{fig:ForcedEntanglement} among others.  At the end of the process, we plot the probability as a function of time with respect to finding the system in all states, with the four relevant states shown in Fig. \ref{fig:ForcedEntanglement}.  These are states where the atoms remain in the ground state and consequently both photons are still in the waveguide modes.  Note that the probabilities do not sum to one except at the beginning and end of the interaction period only because we have neglected to show all states where atoms become excited.  The unusual shape of the curve at the beginning and the end is a result of the different times at which the coupling parameters are turned on and off in accordance with Fig. \ref{fig:Adiabatic}.  Extending the interaction time reveals Rabi oscillations between the two anti-bunched states, shown in Fig. \ref{fig:ForcedEntanglement}b.

We can see that at the end of the process in Fig. \ref{fig:ForcedEntanglement}a, both states where the photons are in separate waveguides are at 50\% probability, so the probability amplitudes are equal in magnitude to $1 / \sqrt{2}$.  The difference in phase between these states were found to be $-\pi / 2$.  This means the final output state is:

\begin{equation}
    | \psi (t_{out}) \rangle = \frac{e^{i\phi}}{\sqrt{2}} \big ( | R \rangle_a | L \rangle_b - i | L \rangle_a | R \rangle_b \big ) | G \rangle_A | G \rangle_B.
    \label{eq:psiOut}
\end{equation}

Note that Eq. (\ref{eq:psiOut}) is equivalent to a Bell state, up to local rotations.

\begin{figure}[ht]
    \centering
    \includegraphics[width=0.45\textwidth]{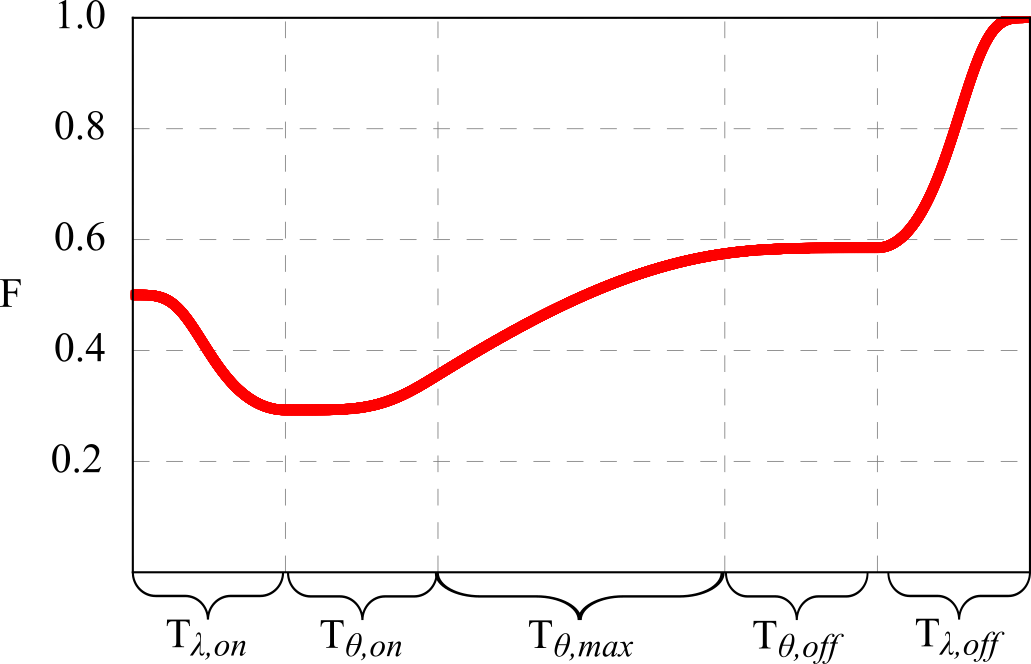}
    \caption{Fidelity as a function of time, for the situation depicted in Fig. \ref{fig:ForcedEntanglement}a.  Initially, fidelity decreases as the absorption coupling is turned on, gradually increasing from then until the end of the simulation.}
    \label{fig:Fidelity}
\end{figure}

In order to confirm that the quantum state is adequately represented by (\ref{eq:psiOut}), we will check it's fidelity, defined $F (\rho,\sigma) \equiv \langle \psi_\rho | \sigma | \psi_\rho \rangle$ \cite{jozsa1994fidelity}.  In our case, $\sigma$ is the density operator describing our simulated system and $| \psi_p \rangle$ will be replaced by the pure state given in (\ref{eq:psiOut}).  The fidelity quantifies how similar the quantum states described by $\rho$ and $\sigma$ are with $F = 1$ signifying identical quantum states.  Note that in our case, $\sigma$ has been obtained by tracing out the absorber component of the quantum system.  We calculate the fidelity as a function of time using identical conditions and time period as Fig. \ref{fig:ForcedEntanglement}a, and plot the results in Fig. \ref{fig:Fidelity}.
We see that The fidelity initially decreases as the absorber coupling is turned on.  As the waveguide coupling is turned on, the fidelity begins to gradually increase, approaching a value of 1 at the end of the simulation.  In other words, the evolution shown in Fig. \ref{fig:ForcedEntanglement}a accurately depicts the growth of an entangled quantum state of polarization.

Similar simulations show that beginning with partial entanglement shows the very same behavior, producing a kind of entanglement distillation.  In the scenario where we suppress bunching, the local maxima and minima of either anti-bunched state probability curve will be the initial probabilities of the two states.  That is, if our input is instead $| \psi (t_{in}) \rangle = \sqrt{3/2} | R \rangle_a | L \rangle_b | G + 1/\sqrt{2} | R \rangle_a | L \rangle_b$, then the least entanglement that can be achieved is this state, or the state with swapped amplitudes.  Interestingly, even in this case the oscillations have precisely the same period as the situation in Fig. \ref{fig:ForcedEntanglement}b.  However, the phase difference between the two anti-bunched states has significant impact on the evolution of the system.  Specifically, it can take noticeably longer for the system to begin the Rabi oscillations, though once they do, the system oscillates as expected.
Notably, since sending a partially entangled state where the two opposed states have 0 phase difference yields smaller and smaller oscillations, it is intuitive that sending in a maximally entangled state should have no oscillations at all, leaving the state in a maximally entangled state throughout the entire process.  Our simulation demonstrated this to be the case as well.

\begin{figure}[ht!]
    \centering
    \includegraphics[width=0.45\textwidth]{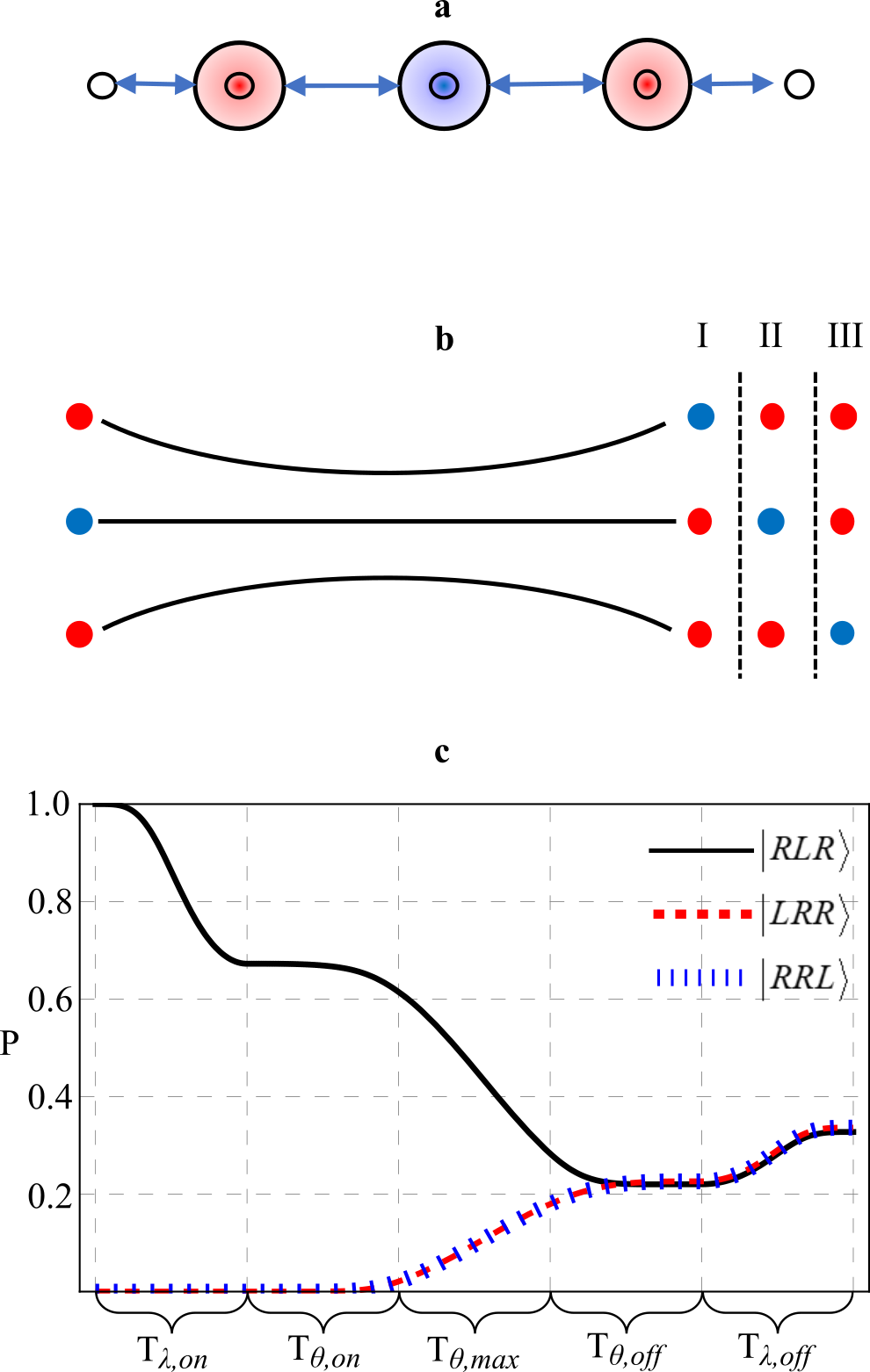}
    \caption{The design and operation of the tripartite entanglement device.  a) Three waveguides placed in a line, coupled to one another with strength $\theta$, and the two end waveguides coupled to absorbers with strength $\lambda$.  b) Three unentangled input photons, two in one polarization state (red) and one in an orthogonal polarization state (blue).  Suppression of bunching produces a quantum superposition of the three possible states where one and only one photon is in each waveguide.  c)  The time evolution of the three relevant states in the superposition.  The red (dashed) and blue (dotted) curves, representing $| LRR \rangle$ and $| RRL \rangle$ respectively, overlap and end up equal to the black (solid) curve, representing $| RLR \rangle$.}
    \label{fig:ZenoW}
\end{figure}

We now extend our analysis to that of three waveguides in a planar configuration for three-photon entanglement generation.
Our setup is nearly identical to that of Fig. \ref{fig:FourLevel}.  We place three waveguides in a line, but only the two waveguides on the end are coupled to diamond-type absorbers, shown in Fig. \ref{fig:ZenoW}.  We input a left polarized photon in the middle path with a right photon in both of the other paths.  Surprisingly, as is discussed in detail in Appendix \ref{app:Three-mode}, we do not couple all three waveguides to absorbers because it actually serves to over restrict the system evolution even in the case of weak coupling.  

The updated Hamiltonian describing this system is changed only slightly with the addition of the new waveguide mode.  The rest Hamiltonian is now given by

\begin{equation}
    \begin{split}
        \hat{H}_0 &= \hbar \omega (\hat{a}_R^\dagger \hat{a}_R + \hat{a}_L^\dagger \hat{a}_L + \hat{b}_R^\dagger \hat{b}_R + \hat{b}_L^\dagger \hat{b}_L + \hat{c}_R^\dagger \hat{c}_R + \hat{c}_L^\dagger \hat{c}_L)\\
        & + \hbar \omega_0 (\hat{A}_3 + \hat{B}_3 + \hat{C}_3)
    \label{eq:RestHam3}
    \end{split}
\end{equation}

and the interaction component is

\begin{equation}
   \begin{split}
        \hat{H}_I (t) &= \hbar \theta (\hat{a}_R^\dagger \hat{b}_R + \hat{b}_R^\dagger \hat{a}_R + \hat{a}_L^\dagger \hat{b}_L + \hat{b}_L^\dagger \hat{a}_L)\\
        & + \hbar \theta (\hat{b}_R^\dagger \hat{c}_R + \hat{c}_R^\dagger \hat{b}_R + \hat{b}_L^\dagger \hat{c}_L + \hat{c}_L^\dagger \hat{b}_L)\\
        & + \hbar \lambda (\hat{A}_{GR} \hat{a}_R + \hat{A}_{LE} \hat{a}_R + \hat{A}_{GL} \hat{a}_L + \hat{A}_{RE} \hat{a}_L)\\
        & + \hbar \lambda (\hat{A}_{RG} \hat{a}^\dagger_R + \hat{A}_{EL} \hat{a}^\dagger_R + \hat{A}_{LG} \hat{a}^\dagger_L + \hat{A}_{ER} \hat{a}^\dagger_L)\\
        & + \hbar \lambda (\hat{C}_{GR} \hat{c}_R + \hat{C}_{LE} \hat{c}_R + \hat{C}_{GL} \hat{c}_L + \hat{C}_{RE} \hat{c}_L )\\
        & + \hbar \lambda (\hat{C}_{RG} \hat{c}^\dagger_R + \hat{C}_{EL} \hat{c}^\dagger_R + \hat{C}_{LG} \hat{c}^\dagger_L + \hat{C}_{ER} \hat{c}^\dagger_L ).
   \end{split}
   \label{eq:IntHam3}
\end{equation}

Contrasting (\ref{eq:RestHam3}) and (\ref{eq:IntHam3}) with (\ref{eq:RestHam}) and (\ref{eq:IntHam}) respectively, waveguide b is the middle waveguide with no absorbers present and waveguide-absorber system b/B in (\ref{eq:RestHam}) and (\ref{eq:IntHam}) are now relabelled to c/C.

Under these conditions where we have a strong Zeno effect acting on the end modes but not the middle, we can achieve a transformation of the input state $| RLR \rangle$ into the superposition state

\begin{equation}
    \begin{split}
        | \Psi_{out} \rangle = \frac{1}{\sqrt{3}} ( &e^{i \phi_{RRL}} | RRL \rangle + e^{i \phi_{RLR}} | RLR \rangle \\
        &+ e^{i \phi_{LRR}} | LRR \rangle ).
    \end{split}
\end{equation}

To local rotations this is a three photon W-state \cite{PhysRevA.65.032108}.

The time-evolution of the three relevant states' probabilities is shown in Fig. \ref{fig:ZenoW}c.  We see that the input state $| RLR \rangle$ gradually decreases and the other two states $| RRL \rangle$ and $| LRR \rangle$ gradually increase until all three end up equal at $1/3$.

Unlike the two-photon case; however, we note that in order to achieve the maximally entangled tripartite state, the coupling parameter $\lambda$ must be carefully tuned.  If it is too small, the transition rates to the bunched states, and consequently two-photon absorption rates, become significant.  On the other hand, if it is too large, then the probabilty amplitudes for the $| LRR \rangle$ and $| RRL \rangle$ components of the superposition cannot grow large enough to equal the input state.  Like the case where absorbers are in all three paths, the Zeno effect can become too strong to allow for perfectly free swapping of the relevant states.  All of these issues will be elaborated upon in Appendix \ref{app:Two-mode}.

\section{Conclusions}

We have shown that the QZE can generate entangled polarization photons when applied to coupled waveguide systems. We first argued using perturbation theory that bunching terms will be suppressed at a higher rate than desired transitions in our system, suggesting that entangled output states are possible. We then performed a numerical simulation of the underlying Schrodinger equation describing our system and observed the impact of various system parameters on output entanglement quality. Finally, we extended our results to systems of three coupled waveguides in a planar configuration and determined through numerical simulation that W states can be generated using this technique.

The QZE is realized in our system using four-level diamond-type atoms that act as observers. 
The possibility of two-photon absorption in our system functions to "project" our system into a subspace that can evolve, in the ideal case, into highly entangled states.

The results here combine some of the ideas from earlier works on the Zeno effect. Specifically, we design a very similar scheme to that of Ref. \cite{franson2004quantum}, but with two key differences. The first is that the design in \cite{franson2004quantum} suppressed bunching in systems of indistinguishable photons, whereas we consider photons distinguishable by their polarization. Secondly, the suppression in \cite{franson2004quantum} is moderated by absorbers with three-level ladder configurations, but the addition of photon polarization in our system necessitates instead absorbers with a diamond-type configuration.

Furthermore, since the two-photon protocol functions in any superposition of the two anti-bunched terms, our scheme can act as a form of entanglement distillation \cite{PhysRevA.53.2046} without utilizing any local operations or external entangled pairs.  In particular, our scheme in the ideal case can distill partially entangled pure quantum states back into fully entangled states. Similar results are expected to hold for our three-waveguide configuration. Our system offers a new path to generating and distilling polarization-entangled states of two and three qubits. 

\section{Acknowledgements}
ICN acknowledges the support of Dr. James Franson whose initial work inspired this manuscript and under whom the work began.  Further, ICN gratefully acknowledges support from the ASEE E-fellows Postdoctoral Fellowship Program.  Work by TAS was supported in part by the U.S. Department of Energy, Office of Science, National Quantum Information Science Research Centers, Co-design Center for Quantum Advantage (C2QA) under contract number DE-SC0012704.

\section{Disclosures}

The authors declare no conflicts of interest.

\newpage

\appendix

\section{Time-dependent Perturbation Theory}
\label{app:TDPT}

This section discusses the well-known mathematics behind perturbation theory \cite{baym1973lectures} which we rely on throughout the manuscript and is included for clarity.  Beginning with the Schrodinger equation we can transform into the interaction picture and treat the interaction component of the Hamiltonian as a perturbation on the system.  An iterative integration of the Schrodinger equation shows that a system beginning in state $| \psi (0) \rangle$ will evolve into some final state $| f \rangle$.  The probability amplitude of this final state at time $t$ is given by projecting $| \psi (0) \rangle$ onto $| f \rangle$.  Each iteration produces a higher order component of this amplitude.  The first and second order probability amplitudes are given by:

\begin{equation}
    \begin{split}
        \langle f | \psi (t) \rangle^{(1)} & = \frac{1}{i \hbar} \int_0^t dt' \langle f | \hat{H}_I (t') | \psi (0) \rangle,\\
        \langle f | \psi (t) \rangle^{(2)} & = \frac{1}{(i \hbar)^2} \int_0^t dt' \int_0^{t'} dt'' \sum_m \langle f | \hat{H}_I (t') | m\rangle \times\\
        &\times \langle m | \hat{H}_I (t'') | \psi (0) \rangle.\\
    \end{split}
    \label{eq:NthOrder}
\end{equation}

Higher order waveguide transitions continue with this pattern.  The sum in Eqn. (\ref{eq:NthOrder}) is over all m where $| m \rangle$ are all intermediate states between $| \psi (0) \rangle$ and $| f \rangle$.  We can obtain the transition probability $P_{i \to f}$ from initial state $| i \rangle$ by taking the magnitude squared of (\ref{eq:NthOrder}).

In accordance with the adiabatic approximation of our particular system, we assume that the waveguide-waveguide and waveguide-atom interaction couplings are turned on slowly over time.  This adds the factor $\text{exp}(\eta t)$ to the perturbation, where $1/\eta$ controls the perturbation's growth rate. The first order transition probability $P_{i \to f}$, from which the transition rate is derived, is given by the equation:

\begin{equation}
    P^{(1)}_{i \to f} = \frac{e^{2 \eta t}}{(E_f - E_i)^2 + (\eta \hbar)^2} | \langle f \big| \hat{H}_I \big| i \rangle |^2.
    \label{eq:ProbFirst}
\end{equation}

Differentiating with respect to t and finally taking the limit $\eta \to 0$, we get the transition rate:

\begin{equation}
    \Gamma^{(1)}_{i \to f} = \frac{2 \pi}{\hbar} | \langle f \big| \hat{H}_I \big| i \rangle |^2 \delta (E_f - E_i).
    \label{eq:GoldFirst}
\end{equation}

The second order transition probability is given by:

\begin{equation}
    P^{(2)}_{i \to f} = \frac{t^2}{\hbar^2} \Big| \sum_m \frac{\langle f | \hat{H}_I | m \rangle \langle m | \hat{H}_I | i \rangle}{E_i - E_m} \Big|^2
    \label{eq:ProbSecond}
\end{equation}

which we can differentiate to get the second order version of the Golden Rule, or the transition rate:

\begin{equation}
     \Gamma^{(2)}_{i \to f} = \frac{2 t}{\hbar^2} \Big| \sum_m \frac{\langle f \big| \hat{H}_I \big| m \rangle \langle m \big| \hat{H}_I \big| i \rangle}{E_i - E_m} \Big|^2.
     \label{eq:GoldSecond}
\end{equation}

The fact that the probability of transition is proportional to $t^2$ is the result of the initial and final states being degenerate.  Higher order terms which we have neglected prevent the overall transition time from ever exceeding 1.

\section{Two-mode Propagation}
\label{app:Two-mode}

Bunching suppression in the two-waveguide system can be achieved with alternative configurations to that described in the main text.
For example, suppression can be achieved with V-type absorbers or with absorbers in only a single waveguide mode.
That is, if only one path is coupled to an absorbing medium, suppression can be seen.  
This is a somewhat surprising result, as one might intuitively think that if one of the two modes is not being observed, that mode at least should see the corresponding bunched state arise in the superposition.  This can be verified by examining a system featuring two-level absorbers in only one of two modes, which suppresses a single photon from swapping waveguides with surprising strength even when far off-resonance.

In this section, we will show that there is in fact more than one Zeno effect occurring in our system, specifically from the off-resonant middle levels which yields a stronger Zeno effect than otherwise would be expected.  First, we will revisit the diamond-type atom, by placing the absorbers in only one mode.  We will follow that up by showing the results for V-type absorbers lying in both modes as well as only one mode.  We will end this section by looking at single-photon propagation in the presence of two-level absorbers in the starting mode.

\subsection{Diamond-type absorbers}
\label{app:SingleDiamond}

\begin{figure}[t!]
    \centering
    \includegraphics[width=0.45\textwidth]{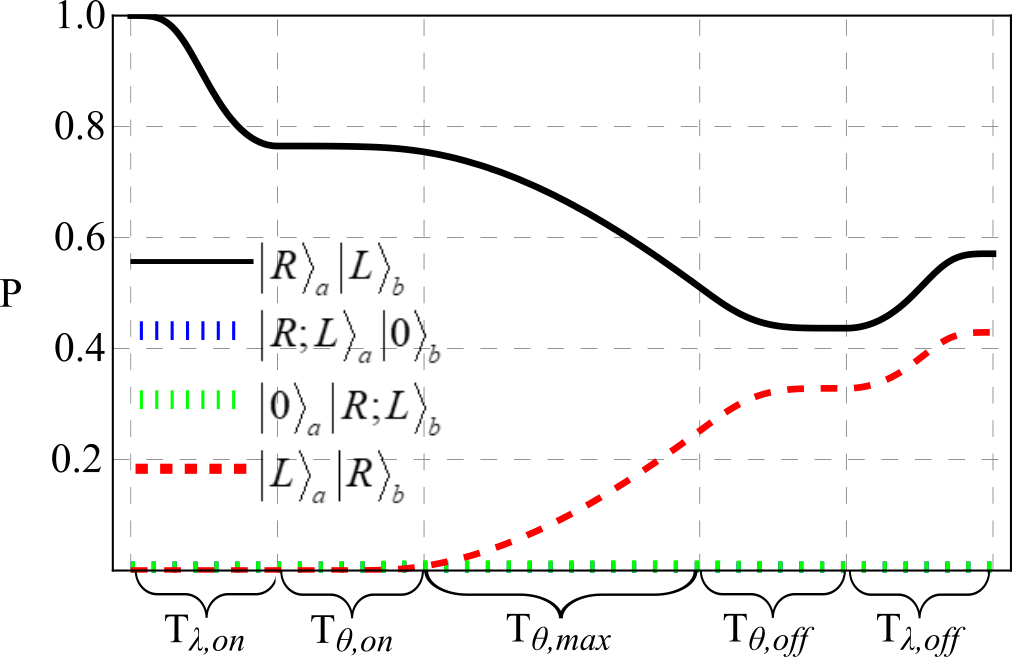}
    \caption{Time-evolution for the two-waveguide system with an absorbing medium coupled to only one of the two modes.  All other conditions are identical to Fig. \ref{fig:ForcedEntanglement}a in the main text.}
    \label{fig:SingleDiamond}
\end{figure}

Under the same conditions as in section 3.1, we remove the absorbing medium from one of the two paths and repeat the simulation.  The time-evolution for this case is shown in Fig. \ref{fig:SingleDiamond}.

What we see in this situation is an evolution of the system nearly identical to the case where both waveguides are separately coupled to identical absorbing mediums, as in the main text.  That is, the probability for the input state (black/solid curve) gradually decreases while the probability for the flipped state (red/dashed curve) gradually increases.  Furthermore, neither of the bunched states (blue and green dotted curves) achieve any significant probability amplitude, indicating successful suppression of those states in the system's evolution.

The only significant difference in the operation of this version of the setup is that the output superposition is not balanced between the two anti-bunched states, indicating the state is not maximally entangled.  Recall that all parameters here are identical to the case in section 3.1, including the time over which the system evolves.  Extending the interaction time reveals that maximal entanglement is achievable.
In other words, we can achieve the same results as in the main text while observing only one out of the two waveguides, but it requires more time over which the process needs to take place.  This is not entirely due to the Zeno effect from the two-photon absorption rate, but also because of the middle levels.  The middle levels produce a second weaker Zeno effect projecting onto single photon states.  Ultimately, the result comes from certain symmetries which are present in the quantum system.  Since the middle levels are equal in energy, the interaction between either left or right photons is identical.  This minor symmetry allows the system to not completely distinguish between the two photon states, and yet simultaneously keep the system anti-bunched as the interaction changes dramatically if the absorbers find a bunched photon state or a vaccuum state.  As we will see, however, bunching suppression from the presence of the excited state is by far the dominant Zeno effect permitting this result.

\subsection{V-type absorbers}
\label{app:V-type}

\begin{figure}[b]
    \centering
    \includegraphics[width=0.3\textwidth]{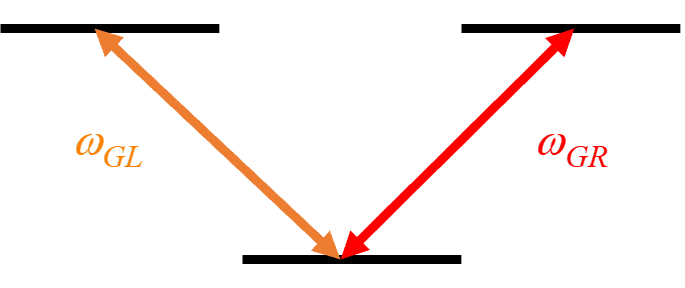}
    \caption{Simple picture for a V-type atom.  One ground state with available transitions to two separate excited states, one resonant with left and one with right circularly polarized light.}
    \label{fig:VType}
\end{figure}

Surprisingly, an anti-bunching Zeno effect is even realizable with V-type absorbers.  An example V-type medium is shown in Fig. \ref{fig:VType}.  Here we assume the same conditions as in the main text and previous section, varying only the interaction time for the process, as well as simulations with absorbers in both paths and absorbers in only one path.  This means that despite having V-type rather than diamond-type structure, we will continue with the same detunings from resonance as we did for the diamond-type medium for the right and left absorptions.

\begin{figure}[ht]
    \centering
    \includegraphics[width=0.45\textwidth]{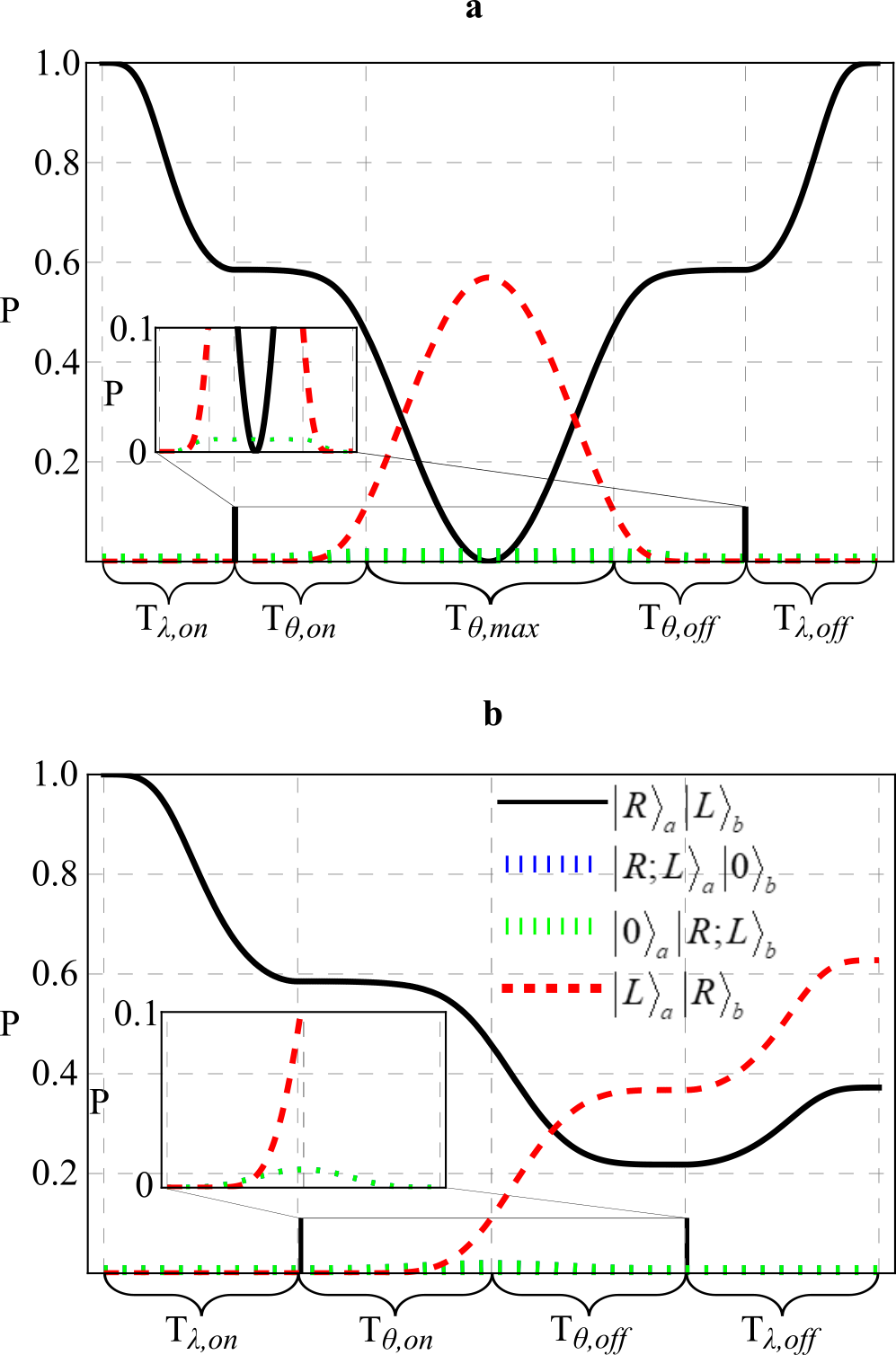}
    \caption{a) Time-evolution for the system with two V-type atoms replacing the diamond-type atoms.  All other conditions (coupling parameters, interaction time) are identical to the situation in Fig. \ref{fig:ForcedEntanglement}a.  The inset plot is a reproduction of the large plot but zoomed in along the y-axis.  b)  A shortened interaction otherwise under the same conditions as a).}
    \label{fig:VForced}
\end{figure}

We begin by simulating the process with absorbers coupled to both waveguide paths.  Fig. \ref{fig:VForced} displays plots showing the time-evolution of this system.  These interactions take place over the same time span as Fig. \ref{fig:ForcedEntanglement}a and Fig. \ref{fig:SingleDiamond}, yet the system has undergone a full Rabi oscillation, that is, the output is identical to the input.  This suggests that achieving an entangled state with V-type absorbers is not only possible, but actually is much faster as well.

The reason for this is the absence of the excited state in the diamond-type atom.  With V-type rather than diamond-type, the medium does not specifically suppress bunching.  Zooming in along the y-axis, shown in the inset plot of Fig. \ref{fig:VForced}, shows that while the bunched state amplitudes remain fairly small, they are not negligible.  The lack of a Zeno effect targeting specifically bunched states allows for more freedom of movement for the photon pair, allowing them to bunch naturally means the second-order process will be faster too.  As suggested in the previous section, the second Zeno effect produced by the left and right levels of the absorber must be present, otherwise the bunched state amplitudes would oscillate alongside the anti-bunched state amplitudes.  What is surprising is that the net effect of the process still yields no bunched states at all.

This holds true even if we shorten the time span of the interaction significantly.  Compare the two plots in Fig. \ref{fig:VForced}.  The conditions for both situations are identical, but the time span for the bottom plot is significantly shorter than the top plot.  Looking at the inset of the top plot, we might suspect that shortening the timespan would leave the bunched state amplitude fairly large, but we see in the inset of the lower plot that it does not.  Notably, in both of these cases, the bunched states are only nonzero while the waveguide modes are coupled together, i.e. when $\theta \neq 0$.

\begin{figure}[t]
    \centering
    \includegraphics[width=0.45\textwidth]{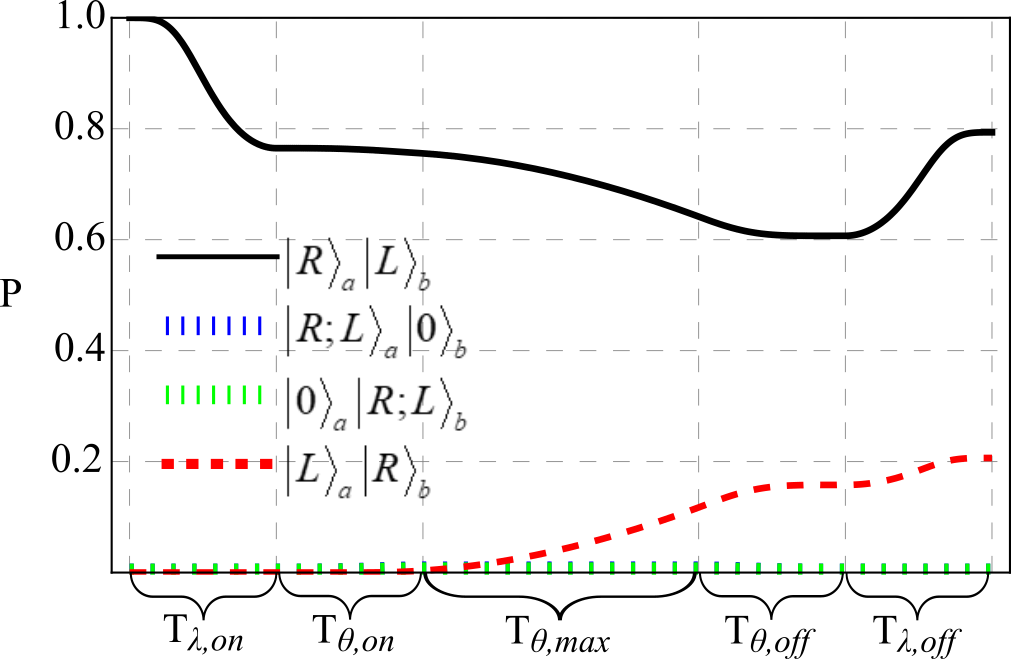}
    \caption{Identical conditions to Fig. \ref{fig:VForced}, but the V-type absorbers are only in one of the two paths.}
    \label{fig:VForcedOne}
\end{figure}

As a final test, we place V-type absorbers in only one of the two paths, just as we did with the diamond-type in section A.1.  Fig. \ref{fig:VForcedOne} shows the results, and we see bunching suppression just as before.  There is some growth of the bunched amplitude in the middle, but just as in Fig. \ref{fig:VForced} it disappears before the end.  Contrasting Figs. \ref{fig:SingleDiamond} and \ref{fig:VForcedOne} we note that it not only takes longer for V-type absorbers to reach an entangled state when they are only present in one mode, but it takes longer than the diamond-type absorbers as well.  Since the V-type absorber possesses the same symmetry as the diamond, the V-type can distinguish between a single-photon state and a two-photon state, but not so much between polarizations.  So the absorber projects the waveguide onto a single-photon state, but the overall interaction changes little if the photons swap places.

While the case where absorbers are in both modes favors the V-type absorber in terms of the necessary coupling time to achieve entanglement, the case where absorbers lie in only one of the two paths favors the diamond-type.  This is because the explicit bunching suppression of the diamond-type not present in the V-type.  Of course, it is slower than having V-type in both paths for the same reason as the diamond-type case, specifically the loss of that symmetry.  However, the system is not explicitly suppressed from evolving into the bunched states, which introduces a new asymmetry into the "permitted" states of the system when compared with the two path absorbers case.  That is, when V-types are in both paths, bunched states are entirely symmetric.  When only in one, these states become asymmetric, causing a more restrictive Zeno effect than in Fig. \ref{fig:VForced}.  Ultimately, this slows the evolution down from Fig. \ref{fig:SingleDiamond} since that explicitly cannot evolve into bunched states.  In other words, the system is biased towards evolution through the bunched states before reaching anti-bunched states, but the quantum evolution is changed drastically by moving from an anti-bunched state to a bunched state.  With explicit suppression of bunching, it can only evolve between anti-bunched states, so this asymmetry does not appear in the evolution.

\subsection{Two-level absorbers}
\label{app:Two-level}

In the previous sections, we have described a second Zeno effect produced by the middle levels of the absorbers, evinced primarily by partial suppression of bunched state evolution in the V-type interactions and certain asymmetries produced by removing absorbers from one of two paths.  In this section, we reduce the total system to the case where only one path is coupled to absorbers, and the absorbers are two-level atoms.  Without loss of generality, we chose a two-level atom which absorbs and emits right-polarized light.  We further reduce the number of incident photons to one right-polarized photon entering the very waveguide which is coupled to said absorbers.  This system removes all relevant symmetries so that we can specifically investigate the relationship between the detuning of the middle levels and the Zeno effect they produce.  In particular, this setup will examine how likely the single photon is to swap waveguides depending on the detuning since the two-level absorber can only project onto or out of a single-photon state.  We will see that the detuning must be much larger than expected before this second Zeno effect has a negligible effect on the system.

Precisely, the relationship between the detuning $\Delta$ and the maximum probability that the photon be found in the second waveguide at the end  will be compared.  All other variables-interaction time, absorber coupling, and waveguide coupling-were held constant, only varying $\omega_R$ for each result.  By "maximum", we mean the largest probability that is achievable, as it is also time-dependent due to Rabi oscillations.  We note that the frequency of oscillation between modes was entirely unchanged by varying the resonant frequency of the absorbers, only the amplitude was changed.  Here, we continue holding $\theta = 1/164$ as in section 3.1, and choose a fairly strong $\lambda = 1/5$, a ratio of $\lambda / \theta = 32.8$.  This is well within the necessary range for bunching suppression to take place in our earlier simulations.

\begin{figure}[t]
    \centering
    \includegraphics[width=0.45\textwidth]{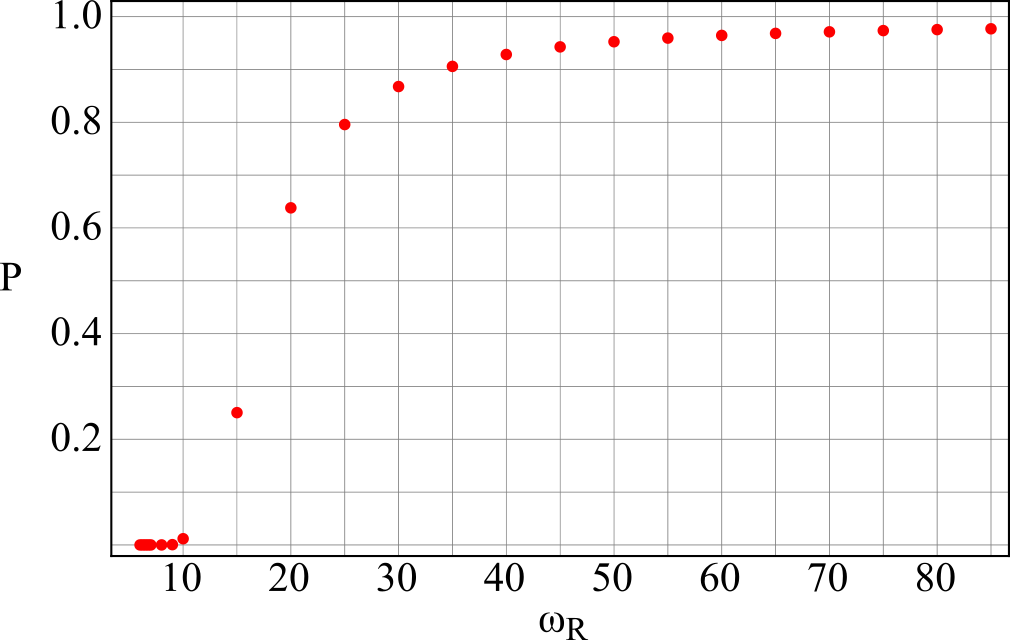}
    \caption{A plot showing the relationship between the resonant frequency $\omega_R = \omega + \Delta \omega_R$ and the maximum probability that a single photon will be found in the second mode.}
    \label{fig:TwoLevel}
\end{figure}

Recall that the detuning is defined by $\omega_R = \omega + \Delta \omega_R$ where $\omega_R$ is the resonant frequency of the absorber and $\omega = 1$ is the frequency of the incident photon.  Fig. \ref{fig:TwoLevel} shows the relationship between the resonant frequency $\omega_R$ and the maximum probability of swapping modes.  $\omega_R$ was varied between 1 and 80 corresponding to a detuning of 0 to 79 or 0\% to 8000\%.  Even for detunings up to 3000\%, we can see that the photon remains in the initial wavguide with 10\%, asymptotically approaching 0\% as we increase the detuning.  Using a smaller $\lambda$ would of course improve this, but may be surprising nonetheles.

In truth, this result should have been expected.  Using perturbation theory, and ignoring the adiabatic approximation for the time being, we can calculate the transition probabilities for absorption $P_\lambda$ and swapping $P_\theta$ separately.  Without the adiabatic approximation, we find the transition probability from simply using eqn. (\ref{eq:WaveguideTrans}) with $\hat{H}_I (t) = \hbar \lambda \exp{(i \Delta \omega_R t)}$ for $P_\lambda$.  $P_\theta$ was already found in section 2.2.  They are

\begin{equation}
    P_\lambda = 4 \lambda^2 \Big [ \frac{\sin{\frac{\Delta \omega_R}{2}t}}{\Delta \omega_R / 2} \Big ]^2
    \label{eq:Two-levelAbsorb}
\end{equation}

and

\begin{equation}
    P_\theta = \theta^2 t^2.
    \label{eq:Two-levelSwap}
\end{equation}

If we take the Taylor expansion of the sine function in (\ref{eq:Two-levelAbsorb}), it becomes clear that

\begin{equation}
    P_\lambda \approx 4 \lambda^2 t^2 \Big [ 1 - \frac{1}{3!} \Big ( \frac{\Delta \omega_R}{2} t \Big )^2 \Big ]
    \label{eq:AbsorbTaylorExpand}
\end{equation}

and the ratio $P_\lambda / P_\theta$ is

\begin{equation}
    \frac{P_\lambda}{P_\theta} = 4 \frac{\lambda^2}{\theta^2} \Big [ \frac{\sin{\frac{\Delta \omega_R}{2}t}}{\Delta \omega_R t / 2} \Big ]^2 \approx 4 \frac{\lambda^2}{\theta^2} \Big [1 - \frac{1}{3!} \Big ( \frac{\Delta \omega_R}{2} t \Big )^2  \Big ].
    \label{eq:ProbRatio}
\end{equation}

We can see that the detuning actually has no effect up to first order, meaning that it \emph{must} be quite large indeed for there to be no noticeable impact on the quantum system by the absorbers.  Instead, we see a Zeno effect keeping the photon from mode swapping, nearly entirely, even up to a detuning of 500\%.  This further corroborates the "weak" Zeno effect which remains present in all cases of this investigation, i.e. two-level, V-type, and diamond-type.

\section{Three-mode Propagation with Three Absorbers}
\label{app:Three-mode}

Here we show the numerical results for the evolution of our input state $| RLR \rangle$ when all three waveguides are coupled to diamond-type absorbers.  We will see that even in the case of relatively small atomic coupling $\lambda$, the suppression is too strong for any transitions between waveguides to occur, even higher order transitions.

\begin{figure}[t]
    \centering
    \includegraphics[width=0.45\textwidth]{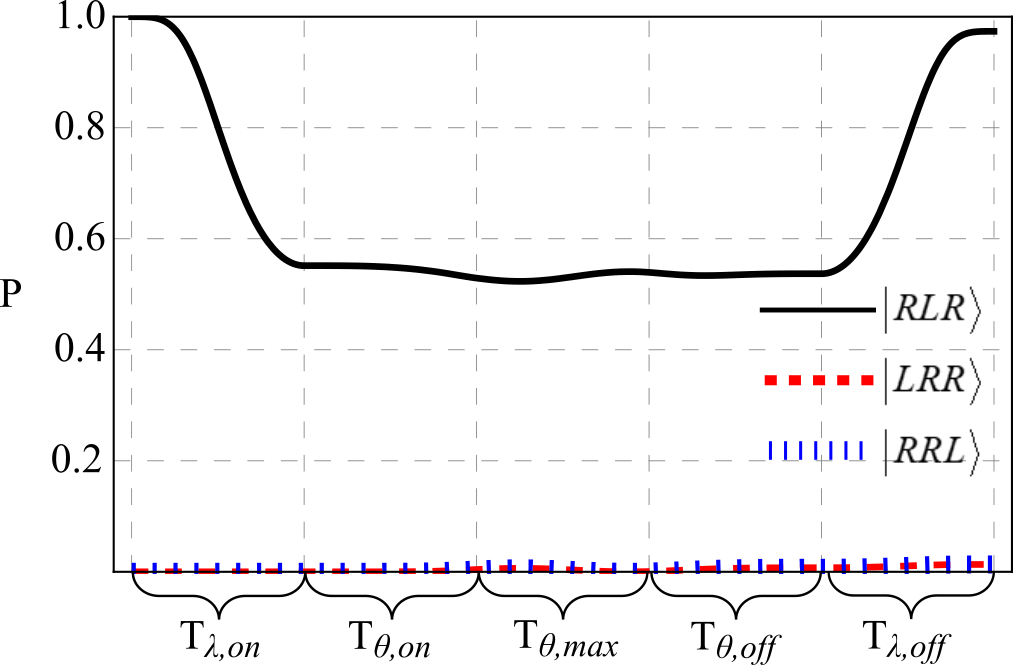}
    \caption{The time-evolution of a three-photon state with one left photon incident in the middle waveguide and one right photon incident in each of the two end waveguides.  The black (solid) curve is the initial state $| RLR \rangle$, while the red (dashed) and blue (dotted) curves are the other two states.}
    \label{fig:WSuppression}
\end{figure}

In Fig. \ref{fig:WSuppression}, we see the time-evolution of the probabilities for all three cases where one photon is in each waveguide, one left and two right, that is, the three states composing a W-state.  There is zero growth of either $| RRL \rangle$ or $| LRR \rangle$, leaving us with, in this particular simulation, a 98\% probability of finding the state in the input case.  Stronger coupling parameters $\lambda$ bring this probability increasingly closer to 100\% while this and weaker coupling parameters leave the system with significant rates of bunching of the two right polarized photons, and consequently two-photon absorption.  In short, designing the device with absorbers in all three paths serves to simply preserve the input state.

\bibliography{bibliography}

\end{document}